\documentclass{mem}
\usepackage{natbib}\usepackage{txfonts}\usepackage{balance}
\usepackage{graphicx}
\usepackage[a4paper]{hyperref}
\idline{75}{1}
\begin{document}
\def\teff{$T\rm_{eff }$}
\def\kms{$\mathrm {km s}^{-1}$}
\def\Omegap{$\Omega_p$}
\def\R1P{R$_1^{\prime}$}
\def\R2P{R$_2^{\prime}$}

\title{
Pattern Speeds and Galaxy Morphology
}

   \subtitle{}

\author{
R. \,Buta\inst{1} 
\and X. \, Zhang\inst{2}
          }

  \offprints{R. Buta}

\institute{
Department of Physics and Astronomy,
University of Alabama, Box 870324, Tuscaloosa, AL 35487
\email{rbuta@bama.ua.edu}
\and
Department of Physics and Astronomy, George Mason University, Fairfax, VA 22030,
\email{xzhang5@gmu.edu}
}

\authorrunning{Buta \& Zhang}

\titlerunning{Pattern Speeds and Galaxy Morphology}

\abstract{
The morphology of a disk galaxy is closely linked
to its kinematic state. This is because
density wave features are likely
made of spontaneously-formed modes which are allowed to arise in the
galactic resonant cavity of a particular basic disk state.
The pattern speed of a density wave is an important parameter that
characterizes the wave and its associated resonances.
Numerical simulations by various authors have enabled us to
interpret some galaxies in terms of high or low pattern speeds.
The potential-density phase-shift method for locating corotation radii
is an effective new tool for utilizing galaxy morphology to
determine the kinematic properties of galaxies. The dynamical
mechanism underlying this association
is also responsible for the secular evolution of galaxies.
We describe recent results from the application of this new method to more than
150 galaxies in the Ohio State University Bright Galaxy 
Survey and other datasets.
}
\maketitle{}

\section{Introduction}

The pattern speed \Omegap\ of a spiral or bar perturbation is an important
physical parameter describing density wave features in galaxies.  Although often
treated as a free parameter in passive numerical simulations, the pattern speed
for a galaxy with a given basic state (described by the radial distribution
of its star/gas surface densities, velocity dispersion, and rotation curve)
is not arbitrary but is determined by the basic state as part of its
spontaneously-formed density wave modal characteristics (Bertin et al. 1989),
although interactions between galaxies can temporarily alter some
of these characteristics.

The pattern speed of a density wave, together with the galaxy rotation
curve, determines the locations of all wave-particle resonances. These resonances
affect the shape of nearby periodic orbits, which can lead to the formation of
 rings (Buta and Combes 1996), and also limit the
extent of the patterns themselves (e.g., Contopoulos 1980; Contopoulos \&
Grosbol 1986). Furthermore,
pattern speed, which determines the location of the
corotation resonance (CR), also influences the manner in which angular
momentum is exchanged with the basic state (Zhang 1998)
when the angular momentum flux is being transported outward by the density wave
(Lynden-Bell and Kalnajs 1972).  This non-uniform angular momentum flux (or
the non-zero radial gradient of the flux) leads to the secular morphological
evolution of galaxies (Zhang 1996, 1998, 1999).
The pattern speed may also influence the lifetime
of a pattern (Merrifield et al. 2006).

To gauge connections between pattern speed and galaxy morphology, one
can (1) directly measure pattern speeds or resonance locations and
connect observed features to resonances; or (2) compare sequences
of numerical simulations with actual galaxies, either through generic
modeling or modeling of a specific galaxy. For example, Garcia-Burillo
et al. (1993), in modeling M51, state that ``The gas response is very
sensitive to the [pattern speed]; we obtain a spiral structure similar
to what is observed only for a narrow range of \Omegap." Included in
(1) are applications of the Tremaine \& Weinberg (1984) continuity
equation method (see other papers in this volume), 
and also our potential-density phase-shift method
(section 3), which taps into a more general connection between galaxy morphology
and galaxy kinematic features.

\section{Pattern Speed and Galactic Rings}

Rings are believed to be direct tracers of wave-particle resonances.
As noted by Buta and Combes (1996),
``rings are a precious tool to measure the pattern speed when the
rotation curve is known." It is worth re-examining models of ringed
galaxies to see how \Omegap\ might affect what we see. The models of
Schwarz (1981), Simkin et al. (1980), and Byrd et al. (1994)
used test-particles to represent gas in analytic, rigidly rotating
bar potentials. Salo et al. (1999) also used gaseous test-particles, but in
a bar potential inferred from a near-infrared image. Rautiainen \& Salo
(2000) used $n$-body simulations to define the bar potential and gaseous
test-particles, and were able to find multiple modes. Lin et al. (2008)
used hydrodynamical models to study the rings in NGC 6782. All of these
models showed that rings are linked to specific orbit resonances in
a bar potential. Other interpretations are provided by Regan \& Teuben (2004),
who describe rings in terms of orbit family transitions rather than
resonances, and by
Romero-Gomez et al. (2006; see also this volume), 
who link rings to invariant manifolds of orbits around the equilibrium
points of the bar.

Pattern speed can have subtle or major effects on a galaxy's structure.
The effects of \Omegap\ for a barred galaxy were explicitly studied by 
Byrd et al. (1994), which confirmed the theoretically expected correlations
between the value of the pattern speed and the type of resonances allowed in a
galaxy. Byrd et al. identified
``pattern speed domains" defined by the presence or absence  of inner
resonances, such as the inner Lindblad resonance (ILR) and the inner 4:1
ultraharmonic resonance (IUHR). High \Omegap\ models had corotation (CR)
and outer Lindblad resonance (OLR), but not ILR or IUHR. Medium \Omegap\
domain models had IUHR, CR, and OLR, but not ILR, while lower \Omegap\
models had all the main resonances. According to the Byrd et
al. models, a galaxy possessing an outer ring or pseudoring but no inner
or nuclear rings would be a candidate for the high \Omegap\ domain,
while a galaxy having inner and outer rings or pseudorings but no
nuclear ring would be a candidate for the medium \Omegap\ domain.
Those galaxies showing three or four possible rings (one or two outer rings, an inner
ring, and a nuclear ring) would be candidates for the low \Omegap\
domain. The domains are nevertheless ambiguous because a galaxy could possess
all the main resonances, and not show all the rings because of a
lack of gas near some resonances.

\begin{figure}[]
\resizebox{\hsize}{!}{\includegraphics[clip=true]{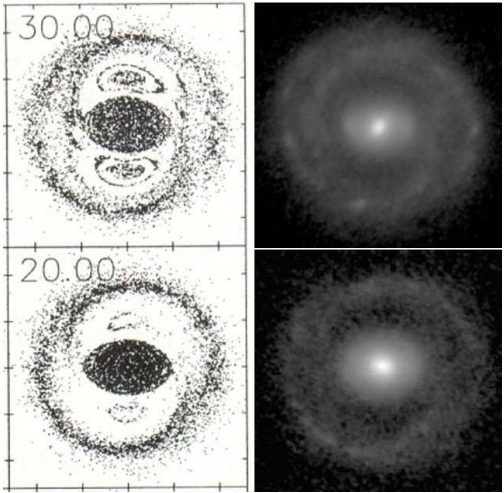}}
\caption{
\footnotesize
These simulated outer pseudoring
patterns at left (top: \Omegap\ = 0.27; bottom: \Omegap\ = 0.22, where
\Omegap\ is dimensionless) occur at the OLR and
were produced in high \Omegap\ domain models by Byrd et al. (1994). 
The number at upper left shows the number of bar rotations. The galaxies
at right are: ESO 509$-$98 (top) and ESO 365$-$35. 
}
\label{high_omegap}
\end{figure}

\begin{figure}[]
\resizebox{\hsize}{!}{\includegraphics[clip=true]{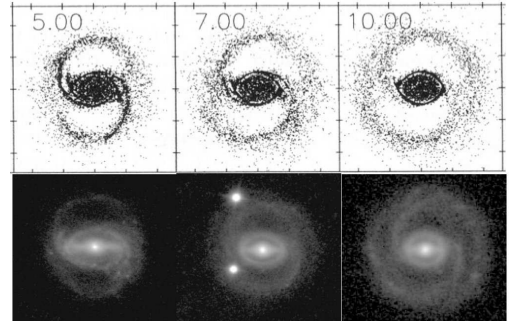}}
\caption{
\footnotesize
A sequence of frames from Byrd et al. (1994) for medium \Omegap\ = 0.10,
covering 5-10 bar rotations.
The comparable galaxies are (left to right): ESO 575$-$47, ESO 426$-$2, and
ESO 577$-$3. All three have conspicuous inner and outer rings/pseudorings,
but lack nuclear star formation based on color index maps from Buta \& Crocker 
(1991).
}
\label{medium_omegap}
\end{figure}

\begin{figure}[]
\resizebox{\hsize}{!}{\includegraphics[clip=true]{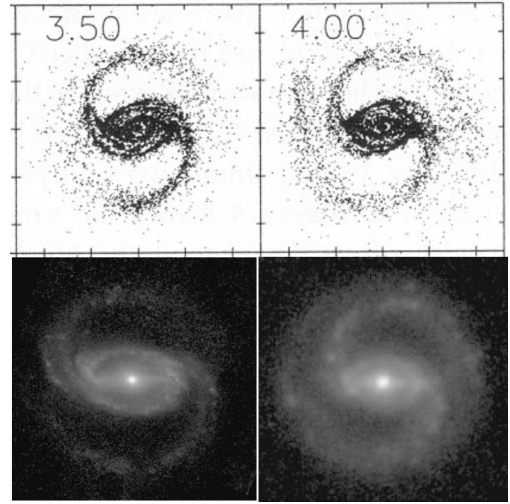}}
\caption{
\footnotesize
Two frames from a low \Omegap\ domain Byrd et al. sequence having \Omegap\ = 0.06, at 3.5 and 4.0 bar rotations.
The model shows inner, outer, and nuclear gas rings/pseudorings. The comparable
galaxies are: ESO 437$-$67 (left) and ESO 325$-$28. The former
has a small blue nuclear ring while the latter has a blue nucleus (Buta \& Crocker 1991).}
\label{low_omegap}
\end{figure}

\begin{figure}[]
\resizebox{\hsize}{!}{\includegraphics[clip=true]{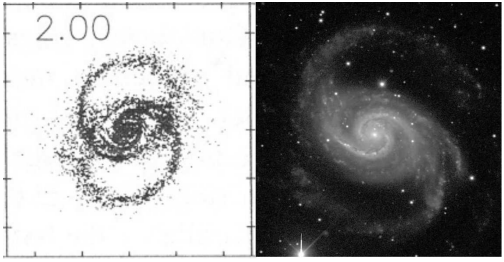}}
\caption{
\footnotesize
A frame from the lowest \Omegap\ sequence of Byrd et al., showing
that a conspicuous spiral develops inside a large oval region, with
an R$_1^{\prime}$ outer pseudoring breaking from its ends. The
galaxy NGC 1566 bears a strong resemblance to this frame.
}
\label{verylow_omegap}
\end{figure}

Figures~\ref{high_omegap}-~\ref{verylow_omegap} compare specific
Byrd et al. simulation frames with observed galaxies mostly from Buta \& Crocker
(1991). These support indirectly the idea of pattern speed domains.
The comparisons highlight not only specific effects of \Omegap\
but also possible effects of morphological evolution (if the sequence is interpreted
as a time frame).
In Figure~\ref{high_omegap}, ESO 509$-$98 [type (R$_1$R$_2^{\prime}$)SB(s)a],
and ESO 365$-$35 [type (R$_2^{\prime}$)S$\underline{{\rm A}}$B(l)0/a]
are two early-type spirals that strongly
resemble high \Omegap\ Byrd et al. models. Both lack an inner ring
or nuclear star formation (Buta \& Crocker 1991), which could
imply the lack of inner resonances.

Figure~\ref{medium_omegap} shows three galaxies
that could be in the medium \Omegap\ domain in that all have
conspicuous inner and outer rings/pseudorings, but lack nuclear rings.
The Byrd et al. sequence shows an R$_1^{\prime}$ outer pseudoring 
evolving to an R$_2^{\prime}$ outer pseudoring, and the three 
illustrated galaxies bear a strong resemblance to the sequence.
Figure~\ref{low_omegap} shows a pair of low \Omegap\ domain frames that again
show an R$_1^{\prime}$ pseudoring evolving to an R$_2^{\prime}$ ring.
The frames include both nuclear rings and inner rings. The two matching
galaxies have similar outer and inner rings, but also from color index
maps are known to have a nuclear ring or blue nucleus, suggesting the
existence of an ILR. Figure~\ref{verylow_omegap} shows a strong resemblance between
the galaxy NGC 1566 and a frame from the lowest \Omegap\ Byrd et al.
model. Instead of a nuclear ring, the model shows a bright inner spiral,
and both galaxy and model have an R$_1^{\prime}$ outer pseudoring.

%In models of ringed galaxy IC 4214, Salo et al. (1999) showed how
%pattern speed affects the appearance and size of rings. Larger rings
%can result from lower values of \Omegap, other things being equal. 
%Two of the simulated patterns
%in Figure~\ref{r1r2p} were discovered in high \Omegap\ models by Schwarz
%(1981), where the term ``high" only means \Omegap\ is high enough to
%avoid inner resonances. These are the familiar R1P and R2P
%subclasses of outer rings and pseudorings.

\begin{figure}[]
\resizebox{\hsize}{!}{\includegraphics[clip=true]{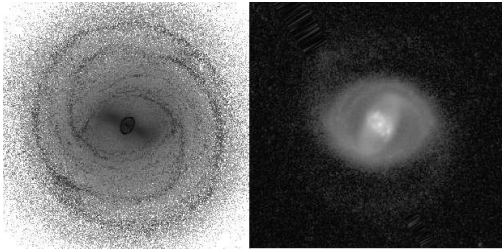}}
\caption{
\footnotesize
The misaligned bar-inner ring galaxy ESO 565$-$11 is a strong candidate
for a multiple \Omegap\ system, according to the models of Rautiainen
\& Salo (2000), one of which is shown at left. 
}
\label{misaligned}
\end{figure}

Although the generic Byrd et al. (1994) models match very well the morphology
of some normal galaxies, the models are limited in allowing only a
single pattern speed.
The Rautiainen \& Salo (2000; see also Rautiainen \& Salo 1999) 
models were more sophisticated in that they allowed
investigation of multiple modes. These authors 
highighted a case where a bar might
appear misaligned with an inner ring because the spiral/ring has a different
pattern speed from the bar. ESO 565$-$11 
very much resembles such a model (Figure~\ref{misaligned}). However, the general alignment
of bars and inner rings (e.g., Buta 1995) would seem to argue that, in many
ringed galaxies at least, the bar and the ring/spiral 
could have the same or a similar pattern speed.

\begin{figure}[]
\resizebox{\hsize}{!}{\includegraphics[clip=true]{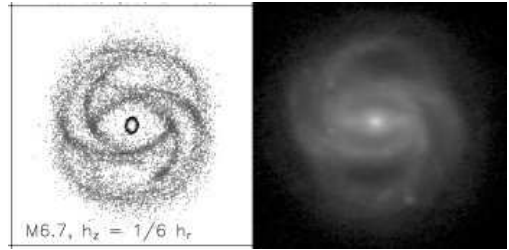}}
\caption{
\footnotesize
A four-armed barred spiral model (left) 
from Rautiainen et al. (2004) as compared
to a $V$-band image of ESO 566$-$24.
}
\label{e566d24}
\end{figure}

Rautiainen et al. (2004) found that an interesting pattern may
develop between the inner and outer 4:1 resonances in a barred
galaxy: a symmetric $m$=4 spiral. Figure~\ref{e566d24} shows a
numerical model where the potential has been derived from an $H$-band image
and evolved mostly with a maximum disk/bulge in the inner regions.
This model shows a four-armed pattern confined between the inner and outer
4:1 resonances that strongly resembles what is seen in ESO 566$-$24. 
A similar conclusion was reached by Treuthardt et al. (2008) for
the spiral morphology of NGC 1433.

This section shows that relatively simple numerical simulations
can account for many aspects of barred galaxy morphology even though
most of these simulations are passive in assuming a static bulge/disk
potential forcing and do not allow the bar to evolve. Also,
except for Rautiainen \& Salo (1999, 2000), these models had assumed
a single pattern speed, and thus are of limited applicability.
The method we describe next allows us to probe galaxy structure
in a more in-depth manner.

\section{The Potential-Density Phase-Shift Method for Locating
Corotation Radii}

\begin{figure}[]
\resizebox{\hsize}{!}{\includegraphics[clip=true]{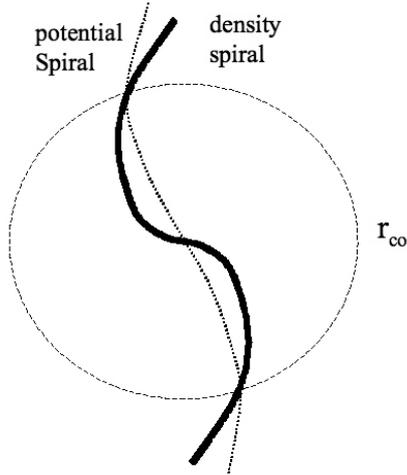}}
\caption{
\footnotesize
Schematic showing the phase difference between potential and density
in a spiral galaxy, and how the difference changes sign across
corotation (labeled $r_{co}$).
}
\label{schematic}
\end{figure}

There are numerous methods for measuring pattern speeds and locating
resonances in galaxies independent of rings (for, example, see other papers in this
volume). The potential-density phase-shift method uses the zero-crossings
of the azimuthal phase difference between the wave perturbation density and the potential
implied by that density, to locate corotation radii in galaxies (Figure ~\ref{schematic}). The method was
conceived by Zhang (1996, 1998, 1999) and first applied to real
galaxies by Zhang \& Buta (2007). The phase difference leads to
a torque applied by the spiral potential on the disk density
(see eq. 1 of Zhang \& Buta 2007).
This torque can lead to a slow secular evolution of the disk surface
density across the entire galactic disk, not just at the main
resonances as advocated by Lynden-Bell \& Kalnajs (1972).
Over a Hubble time, a substantial bulge can be built up by this
process. Note that the method only works if a given pattern
is skewed. Non-skewed patterns will not show phase-shifts.

One requirement for the successful application of this method
is that wave modes in galaxies are quasi-stationary,
which allows the morphological features to track kinematic
characteristics. In such a state, CR
would exactly correspond to positive-to-negative
(P/N) zero crossings in graphs of potential-density
phase differences versus radius (Zhang 1996, 1998). Inside CR,
the density spiral is ahead of the potential spiral,
while outside CR, the density spiral lags the potential spiral 
(Figure~\ref{schematic}). If multiple
patterns with independent \Omegap\ are present, then we will
see multiple P/N crossings. Negative-to-positive (N/P)
crossings appear to delineate
the extent of the modes which may terminate at their OLR.

To calculate the phase-shifts, we use red or near-infrared images to
infer the gravitational potentials, assuming a constant mass-to-light
ratio. Ideally, images in the wavelength range 1.6-4.5$\mu$m are best, 
but even red continuum, Cousins $I$-band, or SDSS $i$-band images are 
useful, depending on Hubble type. When the method is applied to many
galaxies, we note the following trends: (1) multiple P/N crossings
are common, implying nested, multiple patterns; (2) the crossings are
generally not arbitrary but seem closely connected to resonant features;
(3) there is evidence for decoupling of nested patterns, i.e., a bar
and spiral may initially grow together but eventually their pattern
speeds decouple. 

\begin{figure}[]
\resizebox{\hsize}{!}{\includegraphics[clip=true]{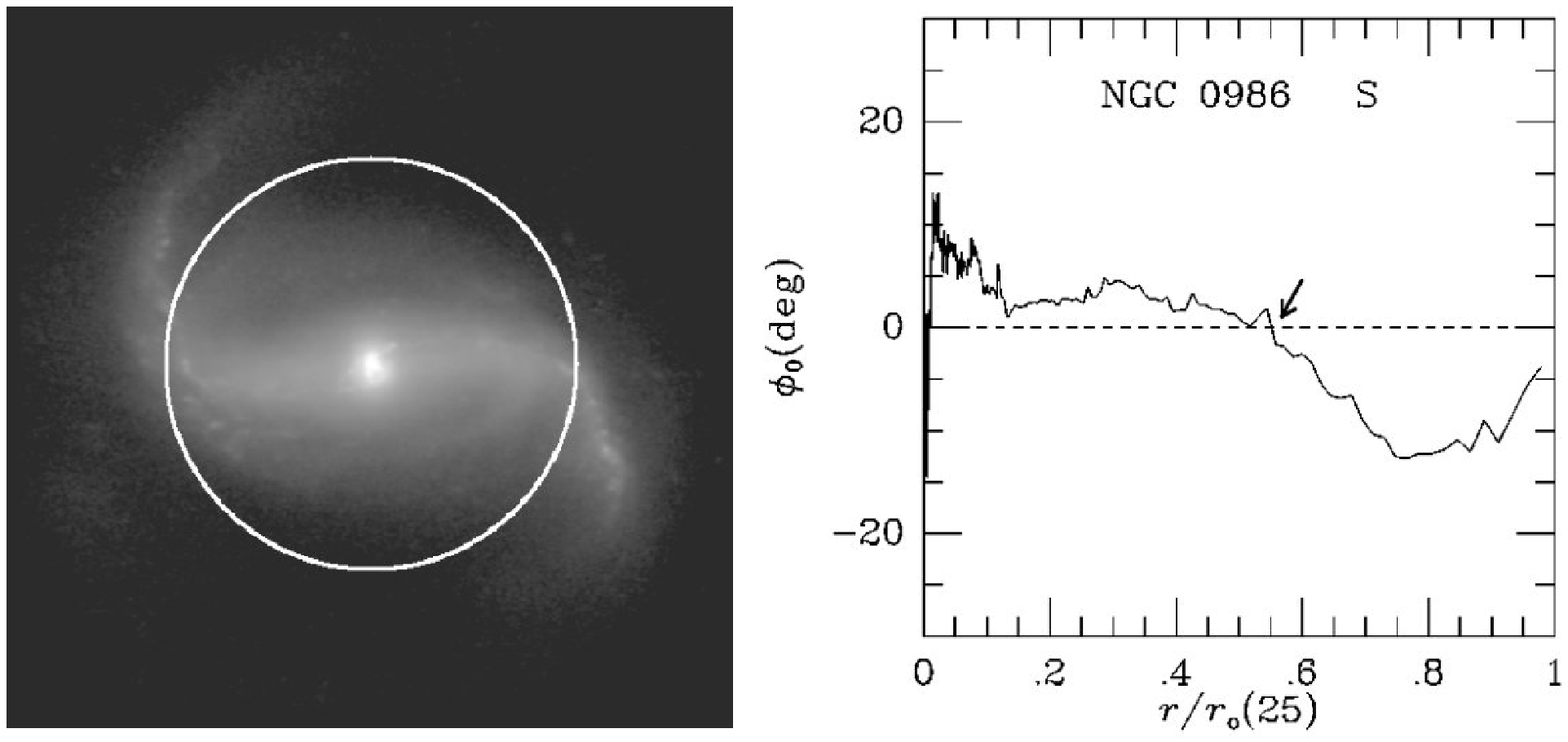}}
\resizebox{\hsize}{!}{\includegraphics[clip=true]{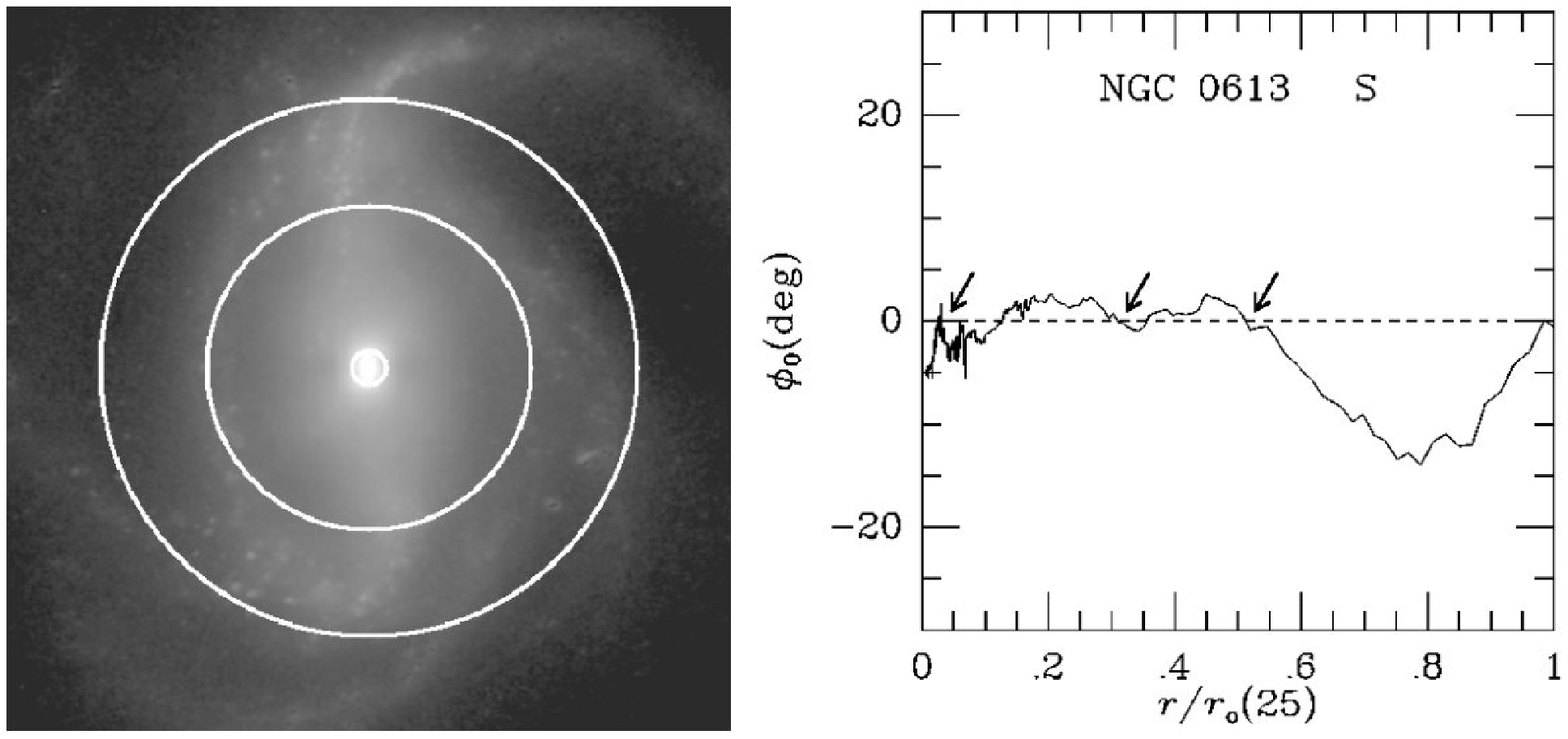}}
\resizebox{\hsize}{!}{\includegraphics[clip=true]{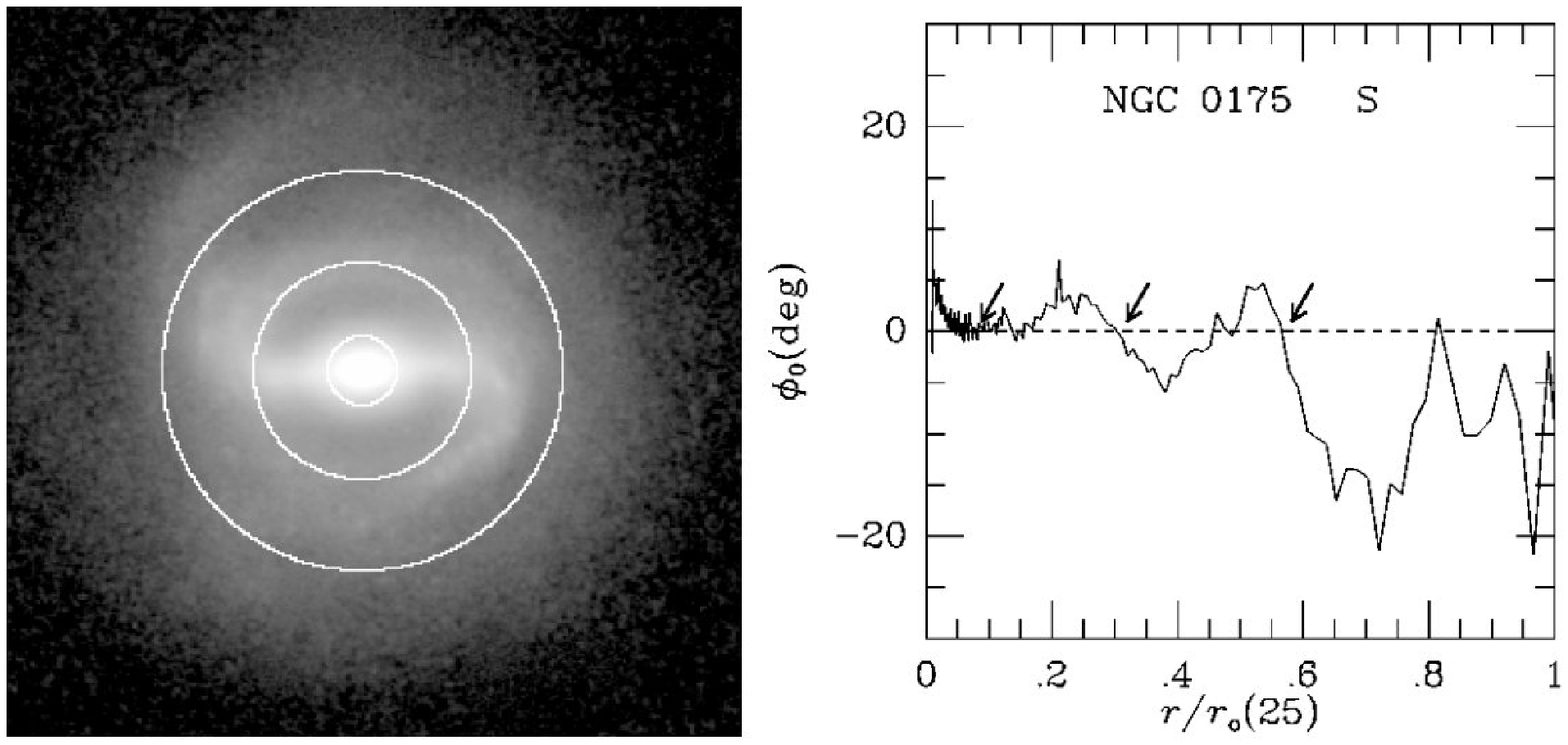}}
\caption{
\footnotesize
$K_s$-band images and phase-shift distributions of three galaxies: (top) NGC 986;
(middle) NGC 613; and (bottom) NGC 175. The arrows on the phase-shift plots
show the main P/N crossings, and hence the implied CR radii. 
The circles superposed on the images show these radii. 
}
\label{phaseshifts}
\end{figure}

The phase-shift distributions $\phi_0(r)$ versus radius $r$, normalized to the
radius, $r_o(25)$, of the $\mu_B$=25.00 mag arcsec$^{-2}$ isophote
corrected for galactic extinction and tilt from RC3 
(de Vaucouleurs et al. 1991),
are shown for three strongly-barred galaxies in Figure~\ref{phaseshifts}.
These are based on $K_s$-band images from Buta et al. (2008).
The first example,
NGC 986, also has a strong spiral breaking from the bar ends and shows
only a single P/N crossing (arrow) in its phase-shift plot. It could be a
genuine case of a ``bar-driven" spiral, where the two features have
grown together and share the same pattern speed. The second example,
NGC 613, shows two crossings in the bar region (plus one near the center), 
which we suggest
indicates that the inner part of the bar is decoupling from the outer
spiral. In the third example, NGC 175, 
the spiral and the bar are fully decoupled
and each has a well-defined P/N crossing. Both NGC 175 and 986 satisfy the
``rule" that the bar extends no further than its CR (Contopoulos 1980).

\begin{figure}[]
\resizebox{\hsize}{!}{\includegraphics[clip=true]{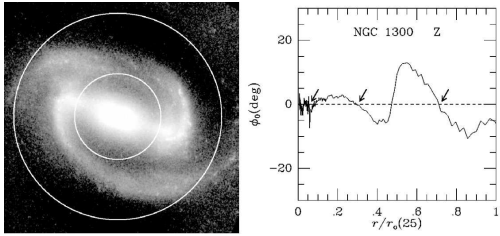}}
\caption{
\footnotesize
$K_s$-band image and derived phase-shift distribution for NGC 1300.
The circles superposed on the image show the two larger derived CR radii.
}
\label{ngc1300}
\end{figure}

Figure~\ref{ngc1300} shows the interesting case of NGC 1300, which has
a well-defined skewed bar and 
two strong P/N crossings corresponding to the smaller and larger
circles on the image. The main P/N crossing associated
with the bar lies inside the prominent bar ansae, while the main outer
crossing is clearly associated with the outer spiral arms. We conclude:
(1) Only the inner portion of the spiral arms associated with
the bar is a  bar-driven spiral; the outer portion appears to break off
at the location of the strong N/P crossing between the two CRs
and is likely to be a distinct mode; and
(2) the prominent bar spills over its CR radius by more than 30\%. We
argued in Zhang \& Buta (2007) that cases where $r(CR)<r(bar)$ are
not necessarily due to image or method noise,
but could be genuine cases violating Contopoulos's finding that a bar
can extend no further than
its own CR, a result which was based on passive orbit analysis in the potential
of a non-skewed bar. 
Support for our finding is that, in order for the SWING mechanism
(Toomre 1981) to amplify spontaneously-formed bars,
a bar cannot terminate cleanly at its CR without a corresponding
wave portion outside CR to receive the transmitted wave. This outer portion
could be in the form of a bar-driven spiral, but could also be in the form
of a slightly-twisted long bar which forms a continuation of the inner bar.

\begin{figure}[]
\resizebox{\hsize}{!}{\includegraphics[clip=true]{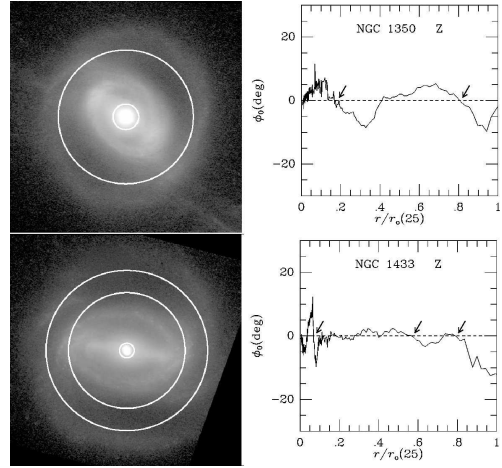}}
\caption{
\footnotesize
Red continuum images (Crocker et al. 1996) and phase-shift distributions
for NGC 1350 (upper frames)
and NGC 1433 (lower frames). The circles show the derived CR radii (arrows).
The phase-shift distribution of NGC 1433 shows additional weak crossings
in the bar region that are likely significant.
}
\label{olrrings}
\end{figure}

Do phase-shift distributions provide any support for the implications
from test-particle and other numerical models of barred spirals, as
outlined in section 2? That is, do outer rings lie outside CR and
inner rings inside CR? The phase-shift distribution of double-ringed
galaxy NGC 1350 
(Figure~\ref{olrrings}, upper frames) shows two P/N crossings corresponding
to the two circles superposed on the image. The smaller circle
seems associated with a pattern inside the bar, while the larger circle
lies between the inner and outer pseudorings. Thus, in this case there
is some support for the findings from the numerical models. The case
of NGC 1433 (Figure~\ref{olrrings}, lower frames) is more complicated.
This galaxy shows two major P/N crossings, one surrounding the ends
of the main bar and inner ring, and the other passing between the outer
pseudoring and the extended oval connected with the inner ring. Interestingly,
the radii of these two inferred CRs are very similar to the radii of the
IUHR and CR derived by Treuthardt et al. (2008) from a
single \Omegap\ numerical simulation of NGC 1433. Our conclusion is that
the bar and oval/spiral of NGC 1433 do not have the same \Omegap.

\begin{figure}[]
\resizebox{\hsize}{!}{\includegraphics[clip=true]{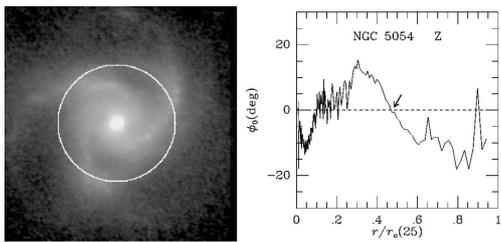}}
\caption{
\footnotesize
$H$-band image of NGC 5054 (Eskridge et al. 2002) and corresponding
phase-shift distribution. The circle superposed on the image corresponds
to the single P/N crossing (arrow).
}
\label{granddesign}
\end{figure}

Do phase-shift distributions support the idea of Contopoulos \&
Grosbol (1986) and Patsis \& Kaufmann (1999) that the main part
of a grand-design spiral extends no further than the 
IUHR, which is typically 2/3 the radius of CR? 
Figure~\ref{granddesign} shows the phase-shift distribution
of the three-armed spiral NGC 5054. Only a single major P/N crossing
is found, and it corresponds to the circle shown on the image.
This circle passes through the middle of the spiral, and suggests
that the pattern most likely extends to the OLR, not the IUHR.
We have found several other cases of a grand design spiral
pattern extending well beyond CR in Zhang \& Buta (2007) and
Buta \& Zhang (2008).

\section{Phase-Shift Results from Analysis of OSUBGS Images}

\begin{figure}[]
\resizebox{\hsize}{!}{\includegraphics[clip=true]{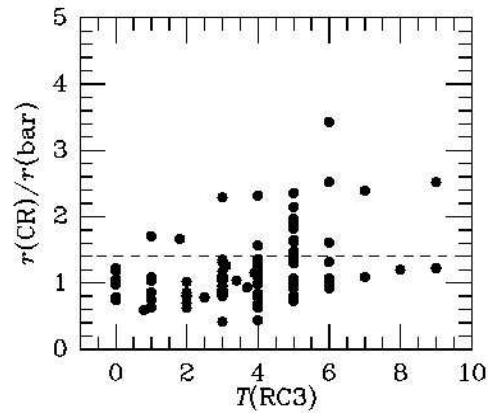}}
\caption{
\footnotesize
Type dependence of the ratio of bar corotation radius to bar major axis
radius, based on phase-shift determined CR radii and published bar radii
from Laurikainen et al. (2004). 
The dashed horizontal line is the division between fast and
slow bars as defined by Debattista \& Sellwood (2000).
}
\label{rcrrbar}
\end{figure}

Buta \& Zhang (2008) have derived phase-shift distributions 
of 153 Ohio State University Bright Galaxy Survey (OSUBGS)
 galaxies based on deprojected $H$-band images 
from the analysis of Laurikainen et al. (2004), who also compiled
estimates of bar radii. With our estimates of bar CR radii from
the phase-shift plots, we have investigated the ratio $\cal R$ =
$r(CR)/r(bar)$, where $r(CR)$ is the bar CR radius judged by P/N
crossings near
the bar radius. $\cal R$ has been linked to the central concentration
of the dark matter halo by Debattista \& Sellwood (2000),
who defined fast bars to have $\cal R$ $\leq$ 1.4 and slow bars to
have $\cal R$ $>$ 1.4. 
Figure~\ref{rcrrbar} shows our estimates of $\cal R$
for 100 galaxies with what Laurikainen et al. called ``Fourier bars,"
which are defined by a constant $m$=2 phase, plotted against the
revised stage index $T$ from RC3. This shows a type-dependence in
the sense that $\cal R$ averages at 1.01$\pm$0.36 for 65 galaxies
of type $T$=4 (Sbc) and earlier, and at 1.48$\pm$0.62 for 35 galaxies
of type $T$=5 (Sc) and later. 

\begin{figure}[]
\resizebox{\hsize}{!}{\includegraphics[clip=true]{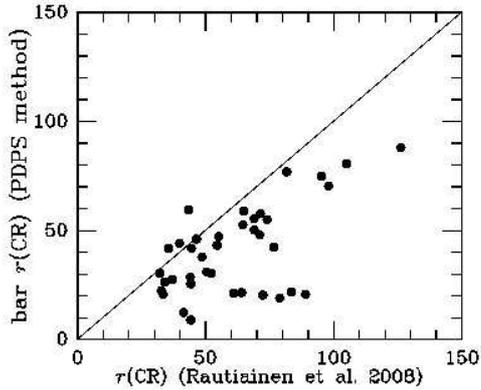}}
\caption{
\footnotesize
A graph of bar CR radii estimated from the potential-density phase-shift
method versus the CR radii estimated for 38 galaxies by Rautiainen et al.
(2008), using the numerical simulation approach.
}
\label{crcomparison}
\end{figure}

How well do CR radii estimated using the potential-density phase-shift
method compare with the results of other approaches? Rautiainen et al.
(2005, 2008) used numerical simulations to estimate CR radii for 38 OSUBGS
galaxies. The idea was to use a near-IR image 
to estimate the potential, and then
evolve a cloud-particle disk in this potential
until the cloud morphology matches the $B$-band appearance of the
spiral arms. Our approach generally gives smaller values of $r(CR)$
than that of Rautiainen et al. (Figure~\ref{crcomparison}), and we found that one of the reasons
for this systematic difference is that the nature of the simulation approach is such that
matching $B$-band spiral morphologies often latches the chosen
solution onto the \Omegap\ of the spiral, and not necessarily the \Omegap\ of the bar.
This conclusion was also arrived at independently by
Rautiainen et al. (2008). 
In spite of the differences shown in Figure~\ref{crcomparison},
Rautiainen et al. (2005) found a similar trend of $\cal R$ with Hubble
type.

\section{Conclusions}

Numerical simulations have been very useful for highlighting
the impact of pattern speed on galaxy structure. From comparisons
with ringed galaxies, there is a suggestion of the existence of
pattern speed domains, where certain major resonances may or may 
not exist. The potential-density phase-shift method suggests
that multiple pattern speeds are common in spiral and barred galaxies.
Both bar-driven and non-bar-driven spirals are detected. The method
also brings attention to the possibility of ``super-fast bars,"
where $r(CR) < r(bar)$.

\begin{acknowledgements}
We thank E. Laurikainen and J. H. Knapen for some of the
images used in this paper.
RB acknowledges the support of NSF Grant AST-0507140.
Funding for the OSUBGS was provided by
NSF grants AST 92-17716 and AST 96-17006, with
additional funding from the Ohio State University.
\end{acknowledgements}

\bibliographystyle{aa}

\end{document}